\documentclass[letterpaper,10pt,journal,twoside]{IEEEtran}

\usepackage[T1]{fontenc}
\usepackage{amsmath,amssymb,amsfonts}
\usepackage{graphicx}
\usepackage{booktabs}
\usepackage{cite}
\usepackage{pgfplots}
\usepgfplotslibrary{groupplots}
\pgfplotsset{compat=1.18}
\usepackage[hidelinks]{hyperref}

\newtheorem{theorem}{Theorem}
\newtheorem{proposition}{Proposition}
\newtheorem{conjecture}{Conjecture}
\newtheorem{assumption}{Assumption}

\newcommand{\C}{\mathbb{C}}
\newcommand{\R}{\mathbb{R}}
\newcommand{\E}{\mathbb{E}}
\newcommand{\Prob}{\mathbb{P}}
\newcommand{\iu}{\mathrm{i}}
\newcommand{\tA}{\tilde{A}^{(d)}}
\newcommand{\Miid}{M_d^{\mathrm{iid}}}

\begin{document}

\title{Minimal Control Time of Large Generic \\ Quantum Systems at Critical Spectral Growth:\\ A Minimum-Spanning-Tree Analysis}

\author{Thomas Chambrion, and Cristian Mendico%
\thanks{This work was supported by the French National Research Agency (ANR) through the QuBiCCS project (ANR-24-CE40-3008-01), by the EIPHI Graduate School (ANR-17-EURE-0002), and by the Bourgogne Franche-Comt\'e Region.}%
\thanks{T. Chambrion is with Institut de Recherche Math\'ematique Avanc\'ee, UMR 7501, Universit\'e de Strasbourg and CNRS, 67000 Strasbourg, France.}%
\thanks{C. Mendico is with Universit\'e Bourgogne Europe, CNRS, IMB UMR 5584, 21000 Dijon, France.}}

\markboth{IEEE Control Systems Letters, Preprint Version}{ Chambrion, Mendico: Minimal Control Time of Large Generic Bilinear Quantum Systems}

\maketitle

\begin{abstract}
The minimal time needed to steer a closed quantum system between two arbitrary states is a fundamental performance limit of quantum control. For single-input bilinear Schr\"odinger systems of dimension $d$ with unbounded controls, this state-to-state control time is bounded from above by the weight $M_d$ of a minimum spanning tree (MST) built from the drift Hamiltonian written in the eigenbasis of the control Hamiltonian. We ask whether the control time of a \emph{generic} system, whose control Hamiltonian has a Haar-distributed eigenbasis, vanishes as $d\to\infty$ in the critical regime of drift eigenvalues growing as $k^{3/2}$. We prove that the finite clusters of the interaction matrix are asymptotically independent Gaussians, analyse the resulting independent surrogate model, for which $M_d \asymp (\ln d)^{-1/2}$ with an explicit finite-size expansion predicting $M_d\sqrt{\ln d}\to 10/3$, and test the exact Haar model by Monte Carlo simulations up to $d=18101$. The exact and surrogate models are statistically indistinguishable in mean at all tested dimensions, and both are consistent with the predicted limit, supporting the conjecture that the control time vanishes, albeit only logarithmically.
\end{abstract}


\section{Introduction}

\IEEEPARstart{H}{ow} fast can a quantum system be driven between two prescribed states? The question is central in quantum technologies, where every operation must be completed well within the coherence time of the device, and it is the subject of the theory of quantum speed limits \cite{CarliniEtAl2006,CanevaPRL2009,HegerfeldtPRL13,DeffnerReview2017,KochEtAl2022}. From the control-theoretic viewpoint it is a minimum-time problem for a bilinear system evolving on a compact Lie group or on a sphere, a problem that the systems and control community has studied intensively for more than two decades.

For two-level systems, the time-optimal synthesis has been obtained with the Pontryagin maximum principle and the geometric theory of optimal syntheses on two-dimensional manifolds, first for cost functionals and gate design on $SU(2)$ \cite{DAlessandroDahleh2001}, then for minimum time with bounded controls and drift \cite{BoscainMason2005,MasonSalmoniBoscainChitour2008,BoscainLong2012,GaronGlaserSugny2013}, for simultaneous control of several qubits \cite{AlbertiniDAlessandro2016}, and for indirect control through an auxiliary system \cite{AlbertiniDAlessandro2014}. Dissipative two-level systems lead to singular extremals and to a much richer synthesis \cite{BonnardChybaSugny2009,LapertEtAl2010}, and time-optimal cooling of Markovian systems has recently been characterised \cite{Malvetti2024}. In higher dimension, time-optimal problems have been solved for three-level and $n$-level systems through $K+P$ (or $K$--$P$) symmetric-space structures and resonance properties of the minimisers \cite{BoscainChambrion2002,Boscain2003,BoscainChambrionCharlot2005,ChangSepulchre2007,AlbertiniDAlessandro2018,AlbertiniDAlessandroSheller2020}, for coupled spin networks via sub-Riemannian geometry \cite{KhanejaBrockettGlaser2001,KhanejaGlaserBrockett2002}, and for quantum transport, parametric oscillators and population transfer \cite{StefanatosLi2014,Stefanatos2017,GrivopoulosBamieh2008}.

These explicit results rely on low dimension or on strong symmetries. Beyond such cases, one must resort to \emph{estimates} of the control time, built on the controllability theory of bilinear quantum systems \cite{AlbertiniDAlessandro2003,Altafini2007,AltafiniTicozzi2012,DongPetersen2010,MirrahimiRouchon2004,BlochBrockettRangan2010,DAlessandroRomano2012,WangSchirmer2010,BoscainChittaroMasonSigalotti2012}. For single-input systems on compact Lie groups, the minimal controllability time is bounded in terms of the drift and of the geometry of the reachable sets \cite{AgrachevChambrion2003,AgrachevChambrion2006}; small-time reachable sets, periodic control laws, energy estimates and Galerkin approximations of bilinear quantum systems were developed in \cite{BCC_CDC2012,BCC_ACC2012,BCC_CDC2013,BCC_TAC2013,BCC_CDC2024}; time-zero controllability and the equivalence between exact and approximate controllability were analysed in \cite{AgrachevBoscainGauthierSigalotti2017,BoscainGauthierRossiSigalotti2015}. In infinite dimension, arbitrarily small control times have been obtained for bilinear Schr\"odinger equations and rotating molecules \cite{duca2025bilinear,beauchard2026logarithmic,ChambrionPozzoli2022,ChambrionPozzoli2023,PozzoliScagliotti2025,LiangEtAl2026,BoscainChambrionMasonSigalottiSugny2009}, a genuinely infinite-dimensional phenomenon, since in any fixed finite dimension the state-to-state control time is positive. Robustness and ensemble aspects, which also involve time-domain performance, are addressed in \cite{KaushikKhaneja2025,ONeilJonckheereSchirmer2024,ONeilJonckheereSchirmer2025,BoscainSigalotti2017,DirrEndeSchulteHerbruggen2019}.

This letter addresses the bridge between the two regimes: for a sequence of finite-dimensional truncations with prescribed drift spectrum and a \emph{generic} control Hamiltonian, does the state-to-state control time vanish as the dimension $d$ grows? Deterministic upper bounds reduce the question to the weight of a minimum spanning tree (MST) of a weighted interaction graph \cite{Harmonique}, and for polynomial spectra $(k^a)_{k\ge1}$ with $a>3/2$ the answer is positive in probability \cite{Random}. The critical case $a=3/2$, where the decay becomes logarithmic and the correlations imposed by the Haar measure matter, is open.

\emph{Contributions.} (i) We formulate the problem for isotropic random control Hamiltonians and state two nested conjectures (Section~\ref{SEC_problem}). (ii) We prove that finite clusters of the interaction matrix are asymptotically independent complex Gaussians, and we analyse the corresponding independent surrogate: $M_d\asymp(\ln d)^{-1/2}$ in probability, with a finite-size expansion predicting $M_d\sqrt{\ln d}\to 10/3$ from below with a $1/\ln d$ correction (Section~\ref{SEC_surrogate}). (iii) Monte Carlo simulations up to $d=18101$ show that the exact Haar model and the surrogate are statistically indistinguishable in mean, and that both are consistent with the predicted limit (Section~\ref{SEC_numerics}).

\section{Problem Formulation}\label{SEC_problem}

Let $\Lambda=(\lambda_k)_{k\ge1}$ be a real sequence with $|\lambda_1|\le|\lambda_2|\le\cdots$. For $d\ge 2$, set $A^{(d)}=\iu\,\mathrm{diag}(\lambda_1,\dots,\lambda_d)\in\mathfrak{u}(d)$ and consider, for $B^{(d)}\in\mathfrak{u}(d)$, the single-input bilinear system
\begin{equation}\label{EQ_system}
\dot\psi(t)=\big(A^{(d)}+u(t)B^{(d)}\big)\psi(t),\qquad \psi(t)\in\mathbb{S}^{2d-1}\subset\C^d,
\end{equation}
with unbounded, piecewise continuous real control $u$. The state-to-state control time $\tau_{SS}^{(d)}$ is the supremum, over all pairs of unit states $(\psi_0,\psi_1)$, of the infimal time needed to steer $\psi_0$ to $\psi_1$.

\begin{proposition}[\cite{Harmonique}]\label{PRO_bounds}
Let $B^{(d)}=PDP^\dagger$ with $D$ diagonal and $P\in U(d)$, and let $\tA=P^\dagger A^{(d)}P$. Let $M_d$ be the weight of the MST of the complete graph on $\{1,\dots,d\}$ with edge weights $W_{jk}=|\tA_{jk}|^{-1}$. Then $\tau_{SS}^{(d)}\ge 2/\sqrt{C_j}$ for every $j$, where $C_j=\sum_{k\ne j}|\tA_{kj}|^2$; if moreover the spectrum of $D$ is rationally independent,
\begin{equation}\label{EQ_upper}
\tau_{SS}^{(d)}\le \tfrac{\pi}{2}\,M_d .
\end{equation}
\end{proposition}

Since $A^{(d)}$ is skew-Hermitian, the weights are symmetric, and both bounds are invariant under $\lambda_k\mapsto\lambda_k+c$. Remarkably, \eqref{EQ_upper} depends on $B^{(d)}$ only through its eigenbasis $P$: with unbounded controls, the eigenvalues of $B^{(d)}$ only need to be rationally independent. Whether $\tau_{SS}^{(d)}\to0$ is therefore decided by the behaviour of $M_d$.

\emph{Generic control Hamiltonians.} To avoid pathological worst cases, we draw $B^{(d)}$ from an \emph{isotropic} ensemble, i.e., a unitarily invariant law on $\mathfrak{u}(d)$ whose eigenvalues have a joint density. By the Weyl integration formula, $P$ is then Haar-distributed on $U(d)$, independently of $D$, and the spectrum of $D$ is almost surely rationally independent. The law of $M_d$ thus depends on $\Lambda$ only. For $\lambda_k=k^a$ with $a>3/2$, typical edge weights scale as $\sqrt{d/v_d}\asymp d^{1/2-a}$ (see \eqref{EQ_vd} below), so that $M_d\asymp d^{3/2-a}\to0$ and $\tau_{SS}^{(d)}\to0$ in probability \cite{Random}. We focus on the critical case $\lambda_k=k^{3/2}$, where the joint law of $\tA$, constrained by orthogonality and trace identities, is intractable.

\begin{conjecture}[General decay]\label{CONJ_general}
For $\lambda_k=k^{3/2}$, $M_d\to0$ in probability as $d\to\infty$; hence $\tau_{SS}^{(d)}\to0$ in probability.
\end{conjecture}

\begin{conjecture}[Precise scaling]\label{CONJ_precise}
There exists $\alpha>0$ such that $\E(M_d)=\alpha(\ln d)^{-1/2}(1+o(1))$.
\end{conjecture}

\section{The Independent Gaussian Surrogate}\label{SEC_surrogate}

Let $v_d=\frac1d\sum_{k=1}^d\lambda_k^2-\big(\frac1d\sum_{k=1}^d\lambda_k\big)^2$ be the empirical spectral variance.

\begin{assumption}[Lyapunov condition]\label{ASS_Lyapunov}
$\frac1d\sum_{k=1}^d\lambda_k^4=o(d\,v_d^2)$ as $d\to\infty$.
\end{assumption}

For $\lambda_k=k^{3/2}$, one has
\begin{equation}\label{EQ_vd}
v_d\sim\tfrac{9}{100}\,d^3,
\end{equation}
and Assumption~\ref{ASS_Lyapunov} holds.

\begin{proposition}\label{PRO_clusters}
Let Assumption~\ref{ASS_Lyapunov} hold and let $P$ be Haar-distributed. Let $S=\{(j_r,k_r)\}_{r=1}^m$ be a fixed finite set of distinct pairs with $j_r<k_r$, and $Z^{(d)}_{jk}=\sqrt{d/v_d}\,\tA_{jk}$. Then $(Z^{(d)}_{j_1k_1},\dots,Z^{(d)}_{j_mk_m})\to\mathcal{CN}(0,I_m)$ in distribution as $d\to\infty$.
\end{proposition}

\begin{IEEEproof}[Sketch of Proof]
\emph{Uncorrelatedness.} For $\Theta=\mathrm{diag}(e^{\iu\theta_1},\dots,e^{\iu\theta_d})$, the Haar law is invariant under $P\mapsto P\Theta$, which maps $\tA_{jk}$ to $e^{\iu(\theta_k-\theta_j)}\tA_{jk}$. Hence $\E[\tA_{jk}\overline{\tA_{j'k'}}]=e^{\iu(\theta_k-\theta_j-\theta_{k'}+\theta_{j'})}\E[\tA_{jk}\overline{\tA_{j'k'}}]$ for all $\theta\in\R^d$, which forces $\E[Z_{jk}\overline{Z_{j'k'}}]=0$ when $(j,k)\neq(j',k')$. With $\E|Z_{jk}|^2=d^2/(d^2-1)$ \cite{Random}, the covariance of the cluster tends to $I_m$.
\emph{Joint normality.} The cluster depends on the $r\le 2m$ columns $P_{\cdot\alpha}$, $\alpha\in\mathcal{K}=\{j_r,k_r\}$. By the Gram--Schmidt representation of Haar columns \cite[Ch.~2]{Meckes_2019}, $P_{\cdot\alpha}=\Pi_{\prec\alpha}G_\alpha/\|\Pi_{\prec\alpha}G_\alpha\|$, with i.i.d.\ $G_\alpha\sim\mathcal{CN}(0,I_d)$ and $\Pi_{\prec\alpha}$ the orthogonal projection onto the complement of the previously generated columns. Writing $\tA_{jk}=\iu\langle\lambda\circ P_{\cdot j},P_{\cdot k}\rangle$, each $Z_{jk}$ equals an envelope factor $e_{jk}^{(d)}=\iu\sqrt{d/v_d}\,\|\Pi_{\prec k}(\lambda\circ P_{\cdot j})\|/\|\Pi_{\prec k}G_k\|$ times $\langle u_{jk},G_k\rangle$, where $u_{jk}$ is a unit vector measurable with respect to the previously generated columns. Conditionally on these, the vector of the $\langle u_{jk},G_k\rangle$ is complex Gaussian. The envelopes converge to $1$ in probability, by the concentration of the spectral variance of Haar vectors \cite{Random} and by $\|\Pi_{\prec k}G_k\|^2/d\to1$. Dominated convergence of characteristic functions yields $Z^{(d)}\to\mathcal{CN}(0,\Sigma)$.
\emph{Identification.} Second moments converge, hence are uniformly integrable, so $\Sigma=\lim\mathrm{Cov}(Z^{(d)})=I_m$.
\end{IEEEproof}

Proposition~\ref{PRO_clusters} motivates replacing $\tA$ by a matrix $G$ with i.i.d.\ strictly upper-triangular entries $G_{jk}\sim\mathcal{CN}(0,\sigma_d^2)$, where $\sigma_d^2=\frac{v_d}{d-1}\cdot\frac{d}{d+1}$ is the exact off-diagonal second moment of $\tA_{jk}$ under the Haar measure. Let $\Miid$ denote the corresponding MST weight. Its edge weights $W_{jk}=|G_{jk}|^{-1}$ are i.i.d.\ with Fr\'echet distribution function
\begin{equation}\label{EQ_Frechet}
F(w)=\Prob(W_{jk}\le w)=\exp\!\big(-1/(\sigma_d^2w^2)\big),
\end{equation}
so they can be sampled directly as $W=(-\sigma_d^2\ln U)^{-1/2}$ with $U$ uniform on $(0,1]$.

\begin{theorem}\label{THE_envelope}
As $d\to\infty$, $\Omega_\Prob\big(d^{3/2}/\sqrt{v_d\ln d}\big)\le\Miid\le O_\Prob\big(d^{3/2}/\sqrt{v_d\ln d}\big)$. For $\lambda_k=k^{3/2}$, $\Miid\asymp_\Prob(\ln d)^{-1/2}$.
\end{theorem}

\begin{IEEEproof}
The MST has $d-1$ edges, each with weight between the global minimum $w_{\min}$ of the $N=d(d-1)/2$ weights and the largest MST weight $w_{\max}$, so $(d-1)w_{\min}\le\Miid\le(d-1)w_{\max}$. By extreme-value theory \cite[Th.~1.5.1]{leadbetter1983extremes}, $NF(w_{\min})=\Theta_\Prob(1)$, which gives $w_{\min}\sim(\sqrt2\,\sigma_d\sqrt{\ln d})^{-1}$. In Kruskal's algorithm, edges are processed by increasing weight, and the last accepted edge is exactly the connectivity threshold of the threshold graph $\mathcal{G}_w=\{e:W_e\le w\}$, which is an Erd\H{o}s--R\'enyi graph $\mathcal{G}(d,F(w))$. Its connectivity threshold $F(w_{\max})\sim\ln d/d$ \cite{erdos1960evolution} gives $w_{\max}\sim(\sigma_d\sqrt{\ln d})^{-1}$. Since $\sigma_d\sim\sqrt{v_d/d}$, the claim follows, and \eqref{EQ_vd} yields the last statement.
\end{IEEEproof}

\emph{Finite-size expansion.} To extract the limit of $y(d)=\Miid\sqrt{\ln d}$ from data, one needs the correction to the \emph{total} weight, not to a single edge. The tail-sum identity gives $\Miid=\int_0^\infty(c(w)-1)\,dw$, where $c(w)$ is the number of connected components of $\mathcal{G}_w$. With $p=F(w)$, \eqref{EQ_Frechet} gives $dw=dp/\big(2\sigma_d\,p\,(-\ln p)^{3/2}\big)$, and the sweep in $p$ splits into three Erd\H{o}s--R\'enyi phases \cite{erdos1960evolution}.

\emph{Forest phase} ($p<1/d$): $\E c=d-\binom d2p+O(1)$, and
\begin{multline*}
\int_0^{1/d}\!\Big(d-\tfrac{d^2}{2}p\Big)\frac{dp}{2\sigma_dp(-\ln p)^{3/2}}\\
=\frac{d}{\sigma_d\sqrt{\ln d}}\Big(1-\frac{1}{4\ln d}+O\big(\tfrac{1}{\ln^2 d}\big)\Big).
\end{multline*}

\emph{Supercritical phase} ($1/d<p<(\ln d+c)/d$): with $t=dp$, the small tree components contribute, by Cayley's formula, $\E[c-1]=d\sum_{k\ge1}\frac{k^{k-2}t^{k-1}e^{-kt}}{k!}(1+o(1))$; freezing the Jacobian at $(\ln d)^{3/2}$, this phase contributes $A_{sc}\,d/(\sigma_d(\ln d)^{3/2})$ with
\begin{equation*}
A_{sc}=\frac12\sum_{k\ge1}\frac{k^{k-2}}{k!}\int_1^\infty t^{k-2}e^{-kt}\,dt\approx0.1389 ,
\end{equation*}
where the isolated vertices ($k=1$) account for $E_1(1)/2\approx0.1097$.

\emph{Connected phase} ($p>(\ln d+c)/d$): $c=1$ and the integrand vanishes.

Summing the three phases yields
\begin{equation}\label{EQ_expansion}
\E\,\Miid=\frac{d}{\sigma_d\sqrt{\ln d}}\Big(1+\frac{A}{\ln d}+o\big(\tfrac{1}{\ln d}\big)\Big),
\end{equation}
with $A=A_{sc}-\tfrac14\approx-0.111$.
For $\lambda_k=k^{3/2}$, \eqref{EQ_vd} gives $d/\sigma_d\to10/3$, hence
\begin{equation}\label{EQ_y}
y(d)=\frac{d}{\sigma_d}\Big(1+\frac{A}{\ln d}+o\big(\tfrac{1}{\ln d}\big)\Big)\longrightarrow\alpha=\frac{10}{3},
\end{equation}
from below since $A<0$. The natural finite-size variable is therefore $x(d)=1/\ln d$, and not the single-edge parameter $\ln\ln d/\ln d$ obtained by inverting $F(w_{\max})=\ln d/d$. The phase boundaries are sharp only asymptotically, and the subleading terms neglected above (exact Jacobian, near-critical window $t\approx1$) are of relative order $1/\ln d$ with respect to $A$, so $A$ is a limit coefficient and its finite-$d$ effective value may differ.

\section{Numerical Investigation}\label{SEC_numerics}

\emph{Protocol.} For the exact model, $P$ is sampled from the Haar measure by QR decomposition of a complex Ginibre matrix with phase correction ($R$ with positive diagonal), $\tA=P^\dagger A^{(d)}P$ is formed with $\lambda_k=k^{3/2}$, and the MST of the weights $|\tA_{jk}|^{-1}$ is computed by Kruskal's algorithm with a union--find structure. The cost is dominated by $O(d^3)$ dense linear algebra. For the surrogate, the $d(d-1)/2$ weights are drawn directly from \eqref{EQ_Frechet} by inverse transform sampling, which reduces the cost to $O(d^2\ln d)$. We used $16$ dimensions from $d=100$ to $d=18101$ (geometric ratio $\sqrt2$), with $N=2000$ realisations per dimension, except for the exact model at $d\in\{9051,12800,18101\}$, where $N\in\{488,500,200\}$. Summary statistics are reported in Table~\ref{TAB_stats}.

\begin{table}[t]
\centering
\caption{Sample means and standard deviations of the MST weight for the surrogate and the exact Haar model, and signal-to-noise ratio of the residual $\Delta M_d=M_d-\Miid$.}\label{TAB_stats}
\footnotesize
\setlength{\tabcolsep}{3.6pt}
\begin{tabular}{rcccccr}
\toprule
 & \multicolumn{2}{c}{Surrogate ($N=2000$)} & \multicolumn{3}{c}{Exact Haar} & \\
\cmidrule(lr){2-3}\cmidrule(lr){4-6}
$d$ & $\langle\Miid\rangle$ & $s\times10^{3}$ & $\langle M_d\rangle$ & $s\times10^{3}$ & $N$ & SNR\\
\midrule
100   & 1.52019 & 20.76 & 1.52018 & 17.50 & 2000 & $-0.02$\\
142   & 1.47117 & 15.69 & 1.47099 & 13.34 & 2000 & $-0.39$\\
200   & 1.42600 & 11.79 & 1.42629 & 10.54 & 2000 & $0.84$\\
283   & 1.38448 & 9.06  & 1.38472 & 8.36  & 2000 & $0.87$\\
400   & 1.34559 & 6.91  & 1.34589 & 6.78  & 2000 & $1.37$\\
566   & 1.30997 & 5.36  & 1.31002 & 5.23  & 2000 & $0.29$\\
800   & 1.27646 & 4.34  & 1.27646 & 4.04  & 2000 & $-0.01$\\
1131  & 1.24535 & 3.22  & 1.24533 & 3.29  & 2000 & $-0.11$\\
1600  & 1.21635 & 2.59  & 1.21629 & 2.55  & 2000 & $-0.76$\\
2263  & 1.18915 & 2.04  & 1.18918 & 1.99  & 2000 & $0.50$\\
3200  & 1.16366 & 1.63  & 1.16367 & 1.54  & 2000 & $0.16$\\
4525  & 1.13972 & 1.26  & 1.13978 & 1.28  & 2000 & $1.66$\\
6400  & 1.11724 & 1.01  & 1.11721 & 1.00  & 2000 & $-0.81$\\
9051  & 1.09595 & 0.80  & 1.09594 & 0.78  & 488  & $-0.30$\\
12800 & 1.07588 & 0.64  & 1.07592 & 0.64  & 500  & $1.25$\\
18101 & 1.05686 & 0.52  & 1.05685 & 0.53  & 200  & $-0.28$\\
\bottomrule
\end{tabular}
\end{table}

\begin{figure}[t]
\centering
\begin{tikzpicture}
\begin{groupplot}[
  group style={group size=3 by 1, horizontal sep=4pt},
  width=0.44\columnwidth, height=0.40\columnwidth,
  scale only axis=false,
  xmin=-4.2, xmax=4.2, ymin=-4.2, ymax=4.2,
  xtick={-3,0,3}, ytick={-3,0,3},
  tick label style={font=\scriptsize}, label style={font=\footnotesize},
  title style={font=\footnotesize, yshift=-4pt},
  grid=major, grid style={gray!25}]
\nextgroupplot[title={$d=100$}, ylabel={Standardised sample quantiles}]
\addplot[only marks, mark=square, mark size=1.4pt, red!80!black] coordinates {(-3.481,-3.036) (-3.175,-2.979) (-3.023,-2.942) (-2.920,-2.928) (-2.841,-2.857) (-2.776,-2.847) (-2.721,-2.776) (-2.674,-2.733) (-2.376,-2.375) (-2.114,-2.198) (-1.956,-2.023) (-1.828,-1.849) (-1.725,-1.744) (-1.638,-1.648) (-1.561,-1.559) (-1.497,-1.515) (-1.434,-1.429) (-1.377,-1.377) (-1.324,-1.334) (-1.274,-1.283) (-1.231,-1.243) (-1.186,-1.191) (-1.144,-1.155) (-1.104,-1.107) (-1.066,-1.075) (-1.031,-1.028) (-0.995,-1.006) (-0.961,-0.980) (-0.928,-0.946) (-0.896,-0.907) (-0.866,-0.883) (-0.835,-0.847) (-0.806,-0.820) (-0.776,-0.783) (-0.748,-0.731) (-0.722,-0.696) (-0.694,-0.673) (-0.667,-0.642) (-0.641,-0.609) (-0.615,-0.592) (-0.591,-0.565) (-0.566,-0.550) (-0.541,-0.521) (-0.517,-0.492) (-0.492,-0.463) (-0.470,-0.442) (-0.446,-0.415) (-0.423,-0.390) (-0.400,-0.368) (-0.377,-0.351) (-0.355,-0.333) (-0.333,-0.310) (-0.310,-0.285) (-0.288,-0.266) (-0.266,-0.243) (-0.245,-0.224) (-0.223,-0.209) (-0.201,-0.187) (-0.180,-0.169) (-0.158,-0.156) (-0.138,-0.140) (-0.116,-0.115) (-0.095,-0.094) (-0.073,-0.080) (-0.052,-0.055) (-0.032,-0.036) (-0.011,-0.006) (0.011,0.012) (0.032,0.037) (0.052,0.060) (0.073,0.080) (0.095,0.104) (0.116,0.124) (0.138,0.151) (0.158,0.175) (0.180,0.189) (0.201,0.209) (0.223,0.233) (0.245,0.248) (0.266,0.272) (0.288,0.299) (0.310,0.316) (0.333,0.338) (0.355,0.360) (0.377,0.380) (0.400,0.400) (0.423,0.421) (0.446,0.433) (0.470,0.457) (0.492,0.483) (0.517,0.515) (0.541,0.533) (0.566,0.554) (0.591,0.576) (0.615,0.597) (0.641,0.619) (0.667,0.645) (0.694,0.665) (0.722,0.704) (0.748,0.727) (0.776,0.764) (0.806,0.789) (0.835,0.819) (0.866,0.846) (0.896,0.872) (0.928,0.907) (0.961,0.941) (0.995,0.974) (1.031,1.021) (1.066,1.043) (1.104,1.095) (1.144,1.145) (1.186,1.188) (1.231,1.241) (1.274,1.264) (1.324,1.331) (1.377,1.372) (1.434,1.431) (1.497,1.492) (1.561,1.559) (1.638,1.643) (1.725,1.710) (1.828,1.828) (1.956,1.925) (2.114,2.118) (2.376,2.470) (2.674,2.773) (2.721,2.834) (2.776,2.837) (2.841,2.847) (2.920,2.885) (3.023,2.916) (3.175,3.066) (3.481,3.241)};
\addplot[only marks, mark=*, mark size=0.7pt, blue] coordinates {(-3.481,-3.453) (-3.175,-2.845) (-3.023,-2.757) (-2.920,-2.716) (-2.841,-2.673) (-2.776,-2.642) (-2.721,-2.622) (-2.674,-2.611) (-2.376,-2.413) (-2.114,-2.173) (-1.956,-1.972) (-1.828,-1.850) (-1.725,-1.721) (-1.638,-1.596) (-1.561,-1.544) (-1.497,-1.486) (-1.434,-1.424) (-1.377,-1.345) (-1.324,-1.292) (-1.274,-1.261) (-1.231,-1.217) (-1.186,-1.177) (-1.144,-1.139) (-1.104,-1.095) (-1.066,-1.050) (-1.031,-1.022) (-0.995,-0.992) (-0.961,-0.940) (-0.928,-0.896) (-0.896,-0.870) (-0.866,-0.846) (-0.835,-0.809) (-0.806,-0.791) (-0.776,-0.770) (-0.748,-0.739) (-0.722,-0.722) (-0.694,-0.698) (-0.667,-0.679) (-0.641,-0.640) (-0.615,-0.620) (-0.591,-0.588) (-0.566,-0.568) (-0.541,-0.546) (-0.517,-0.519) (-0.492,-0.489) (-0.470,-0.466) (-0.446,-0.445) (-0.423,-0.421) (-0.400,-0.398) (-0.377,-0.373) (-0.355,-0.352) (-0.333,-0.331) (-0.310,-0.313) (-0.288,-0.296) (-0.266,-0.274) (-0.245,-0.253) (-0.223,-0.233) (-0.201,-0.211) (-0.180,-0.192) (-0.158,-0.170) (-0.138,-0.152) (-0.116,-0.140) (-0.095,-0.119) (-0.073,-0.087) (-0.052,-0.070) (-0.032,-0.055) (-0.011,-0.033) (0.011,-0.014) (0.032,0.016) (0.052,0.039) (0.073,0.064) (0.095,0.089) (0.116,0.110) (0.138,0.129) (0.158,0.152) (0.180,0.175) (0.201,0.192) (0.223,0.211) (0.245,0.238) (0.266,0.254) (0.288,0.287) (0.310,0.306) (0.333,0.332) (0.355,0.359) (0.377,0.377) (0.400,0.400) (0.423,0.419) (0.446,0.443) (0.470,0.464) (0.492,0.492) (0.517,0.509) (0.541,0.548) (0.566,0.569) (0.591,0.592) (0.615,0.611) (0.641,0.632) (0.667,0.658) (0.694,0.671) (0.722,0.711) (0.748,0.738) (0.776,0.768) (0.806,0.793) (0.835,0.821) (0.866,0.857) (0.896,0.880) (0.928,0.923) (0.961,0.958) (0.995,0.991) (1.031,1.012) (1.066,1.054) (1.104,1.087) (1.144,1.132) (1.186,1.176) (1.231,1.215) (1.274,1.268) (1.324,1.321) (1.377,1.358) (1.434,1.398) (1.497,1.473) (1.561,1.541) (1.638,1.624) (1.725,1.733) (1.828,1.863) (1.956,1.980) (2.114,2.193) (2.376,2.537) (2.674,2.701) (2.721,2.768) (2.776,2.807) (2.841,2.826) (2.920,3.002) (3.023,3.099) (3.175,3.449) (3.481,3.676)};
\addplot[black, thin, domain=-4.2:4.2] {x};
\nextgroupplot[title={$d=1600$}, yticklabels={}, xlabel={Standard normal quantiles}]
\addplot[only marks, mark=square, mark size=1.4pt, red!80!black] coordinates {(-3.481,-4.025) (-3.175,-3.680) (-3.023,-3.382) (-2.920,-3.139) (-2.841,-3.017) (-2.776,-2.971) (-2.721,-2.854) (-2.674,-2.808) (-2.376,-2.471) (-2.114,-2.180) (-1.956,-1.946) (-1.828,-1.838) (-1.725,-1.756) (-1.638,-1.663) (-1.561,-1.604) (-1.497,-1.509) (-1.434,-1.435) (-1.377,-1.373) (-1.324,-1.335) (-1.274,-1.282) (-1.231,-1.220) (-1.186,-1.164) (-1.144,-1.131) (-1.104,-1.083) (-1.066,-1.051) (-1.031,-1.018) (-0.995,-0.979) (-0.961,-0.935) (-0.928,-0.909) (-0.896,-0.869) (-0.866,-0.851) (-0.835,-0.828) (-0.806,-0.794) (-0.776,-0.761) (-0.748,-0.732) (-0.722,-0.697) (-0.694,-0.674) (-0.667,-0.654) (-0.641,-0.621) (-0.615,-0.602) (-0.591,-0.574) (-0.566,-0.554) (-0.541,-0.528) (-0.517,-0.511) (-0.492,-0.490) (-0.470,-0.465) (-0.446,-0.439) (-0.423,-0.420) (-0.400,-0.378) (-0.377,-0.356) (-0.355,-0.342) (-0.333,-0.325) (-0.310,-0.304) (-0.288,-0.288) (-0.266,-0.265) (-0.245,-0.241) (-0.223,-0.211) (-0.201,-0.196) (-0.180,-0.179) (-0.158,-0.163) (-0.138,-0.141) (-0.116,-0.115) (-0.095,-0.090) (-0.073,-0.071) (-0.052,-0.037) (-0.032,-0.017) (-0.011,0.001) (0.011,0.017) (0.032,0.035) (0.052,0.056) (0.073,0.075) (0.095,0.100) (0.116,0.136) (0.138,0.154) (0.158,0.172) (0.180,0.187) (0.201,0.209) (0.223,0.229) (0.245,0.257) (0.266,0.273) (0.288,0.297) (0.310,0.324) (0.333,0.344) (0.355,0.365) (0.377,0.387) (0.400,0.400) (0.423,0.419) (0.446,0.445) (0.470,0.468) (0.492,0.497) (0.517,0.529) (0.541,0.553) (0.566,0.585) (0.591,0.606) (0.615,0.636) (0.641,0.655) (0.667,0.677) (0.694,0.706) (0.722,0.739) (0.748,0.759) (0.776,0.786) (0.806,0.815) (0.835,0.838) (0.866,0.873) (0.896,0.899) (0.928,0.928) (0.961,0.956) (0.995,0.993) (1.031,1.025) (1.066,1.067) (1.104,1.095) (1.144,1.140) (1.186,1.179) (1.231,1.237) (1.274,1.268) (1.324,1.324) (1.377,1.365) (1.434,1.423) (1.497,1.479) (1.561,1.515) (1.638,1.572) (1.725,1.715) (1.828,1.850) (1.956,2.015) (2.114,2.122) (2.376,2.294) (2.674,2.496) (2.721,2.509) (2.776,2.524) (2.841,2.557) (2.920,2.595) (3.023,2.613) (3.175,2.708) (3.481,2.729)};
\addplot[only marks, mark=*, mark size=0.7pt, blue] coordinates {(-3.481,-3.388) (-3.175,-3.218) (-3.023,-3.011) (-2.920,-2.990) (-2.841,-2.961) (-2.776,-2.608) (-2.721,-2.604) (-2.674,-2.535) (-2.376,-2.285) (-2.114,-2.064) (-1.956,-1.921) (-1.828,-1.794) (-1.725,-1.714) (-1.638,-1.645) (-1.561,-1.585) (-1.497,-1.525) (-1.434,-1.472) (-1.377,-1.399) (-1.324,-1.325) (-1.274,-1.278) (-1.231,-1.244) (-1.186,-1.193) (-1.144,-1.138) (-1.104,-1.114) (-1.066,-1.085) (-1.031,-1.054) (-0.995,-0.989) (-0.961,-0.924) (-0.928,-0.897) (-0.896,-0.867) (-0.866,-0.836) (-0.835,-0.809) (-0.806,-0.779) (-0.776,-0.752) (-0.748,-0.724) (-0.722,-0.702) (-0.694,-0.681) (-0.667,-0.661) (-0.641,-0.640) (-0.615,-0.620) (-0.591,-0.597) (-0.566,-0.583) (-0.541,-0.555) (-0.517,-0.526) (-0.492,-0.508) (-0.470,-0.487) (-0.446,-0.445) (-0.423,-0.425) (-0.400,-0.409) (-0.377,-0.381) (-0.355,-0.366) (-0.333,-0.338) (-0.310,-0.310) (-0.288,-0.294) (-0.266,-0.276) (-0.245,-0.254) (-0.223,-0.240) (-0.201,-0.222) (-0.180,-0.195) (-0.158,-0.165) (-0.138,-0.144) (-0.116,-0.126) (-0.095,-0.112) (-0.073,-0.094) (-0.052,-0.066) (-0.032,-0.047) (-0.011,-0.021) (0.011,0.003) (0.032,0.021) (0.052,0.039) (0.073,0.059) (0.095,0.091) (0.116,0.119) (0.138,0.133) (0.158,0.144) (0.180,0.159) (0.201,0.181) (0.223,0.201) (0.245,0.223) (0.266,0.238) (0.288,0.266) (0.310,0.284) (0.333,0.302) (0.355,0.317) (0.377,0.334) (0.400,0.355) (0.423,0.377) (0.446,0.408) (0.470,0.426) (0.492,0.452) (0.517,0.474) (0.541,0.515) (0.566,0.554) (0.591,0.585) (0.615,0.613) (0.641,0.635) (0.667,0.661) (0.694,0.681) (0.722,0.710) (0.748,0.741) (0.776,0.771) (0.806,0.802) (0.835,0.834) (0.866,0.876) (0.896,0.916) (0.928,0.968) (0.961,0.999) (0.995,1.046) (1.031,1.068) (1.066,1.108) (1.104,1.148) (1.144,1.179) (1.186,1.221) (1.231,1.255) (1.274,1.286) (1.324,1.344) (1.377,1.404) (1.434,1.462) (1.497,1.553) (1.561,1.625) (1.638,1.681) (1.725,1.797) (1.828,1.906) (1.956,1.970) (2.114,2.085) (2.376,2.329) (2.674,2.584) (2.721,2.652) (2.776,2.687) (2.841,2.701) (2.920,2.727) (3.023,2.858) (3.175,2.997) (3.481,3.051)};
\addplot[black, thin, domain=-4.2:4.2] {x};
\nextgroupplot[title={$d=12800$}, yticklabels={}]
\addplot[only marks, mark=square, mark size=1.4pt, red!80!black] coordinates {(-3.090,-2.703) (-2.748,-2.623) (-2.576,-2.595) (-2.457,-2.299) (-2.366,-2.271) (-2.290,-2.250) (-2.226,-2.237) (-2.170,-2.218) (-2.120,-2.125) (-1.927,-1.968) (-1.812,-1.866) (-1.717,-1.728) (-1.635,-1.528) (-1.563,-1.449) (-1.483,-1.374) (-1.426,-1.339) (-1.372,-1.307) (-1.323,-1.278) (-1.276,-1.229) (-1.221,-1.196) (-1.180,-1.179) (-1.141,-1.120) (-1.103,-1.072) (-1.067,-1.054) (-1.032,-1.033) (-0.990,-0.984) (-0.958,-0.925) (-0.927,-0.900) (-0.896,-0.868) (-0.867,-0.855) (-0.831,-0.825) (-0.803,-0.815) (-0.776,-0.804) (-0.749,-0.791) (-0.722,-0.722) (-0.690,-0.673) (-0.665,-0.652) (-0.640,-0.629) (-0.616,-0.600) (-0.592,-0.582) (-0.562,-0.531) (-0.539,-0.507) (-0.516,-0.492) (-0.493,-0.478) (-0.470,-0.442) (-0.443,-0.426) (-0.421,-0.405) (-0.399,-0.382) (-0.377,-0.360) (-0.356,-0.339) (-0.329,-0.326) (-0.308,-0.317) (-0.287,-0.298) (-0.266,-0.264) (-0.246,-0.238) (-0.225,-0.218) (-0.199,-0.204) (-0.179,-0.190) (-0.159,-0.180) (-0.138,-0.164) (-0.118,-0.156) (-0.093,-0.128) (-0.073,-0.085) (-0.053,-0.070) (-0.033,-0.064) (-0.013,-0.056) (0.013,-0.021) (0.033,-0.014) (0.053,0.016) (0.073,0.025) (0.093,0.038) (0.118,0.049) (0.138,0.061) (0.159,0.082) (0.179,0.109) (0.199,0.154) (0.225,0.177) (0.246,0.206) (0.266,0.226) (0.287,0.243) (0.308,0.277) (0.329,0.286) (0.356,0.299) (0.377,0.316) (0.399,0.351) (0.421,0.365) (0.443,0.377) (0.470,0.401) (0.493,0.425) (0.516,0.447) (0.539,0.470) (0.562,0.526) (0.592,0.556) (0.616,0.566) (0.640,0.611) (0.665,0.634) (0.690,0.649) (0.722,0.672) (0.749,0.700) (0.776,0.714) (0.803,0.728) (0.831,0.747) (0.867,0.765) (0.896,0.787) (0.927,0.912) (0.958,0.945) (0.990,0.960) (1.032,1.045) (1.067,1.113) (1.103,1.189) (1.141,1.225) (1.180,1.272) (1.221,1.293) (1.276,1.341) (1.323,1.397) (1.372,1.422) (1.426,1.455) (1.483,1.519) (1.563,1.664) (1.635,1.692) (1.717,1.767) (1.812,1.900) (1.927,2.027) (2.120,2.187) (2.170,2.245) (2.226,2.424) (2.290,2.441) (2.366,2.592) (2.457,2.666) (2.576,2.799) (2.748,2.829) (3.090,3.119)};
\addplot[only marks, mark=*, mark size=0.7pt, blue] coordinates {(-3.481,-3.194) (-3.175,-2.958) (-3.023,-2.923) (-2.920,-2.861) (-2.841,-2.816) (-2.776,-2.767) (-2.721,-2.633) (-2.674,-2.600) (-2.376,-2.366) (-2.114,-2.140) (-1.956,-1.985) (-1.828,-1.890) (-1.725,-1.730) (-1.638,-1.664) (-1.561,-1.606) (-1.497,-1.519) (-1.434,-1.471) (-1.377,-1.382) (-1.324,-1.339) (-1.274,-1.272) (-1.231,-1.223) (-1.186,-1.182) (-1.144,-1.144) (-1.104,-1.096) (-1.066,-1.050) (-1.031,-1.029) (-0.995,-0.991) (-0.961,-0.955) (-0.928,-0.930) (-0.896,-0.909) (-0.866,-0.880) (-0.835,-0.849) (-0.806,-0.816) (-0.776,-0.776) (-0.748,-0.739) (-0.722,-0.720) (-0.694,-0.703) (-0.667,-0.683) (-0.641,-0.652) (-0.615,-0.620) (-0.591,-0.596) (-0.566,-0.562) (-0.541,-0.535) (-0.517,-0.501) (-0.492,-0.482) (-0.470,-0.448) (-0.446,-0.423) (-0.423,-0.401) (-0.400,-0.375) (-0.377,-0.347) (-0.355,-0.325) (-0.333,-0.302) (-0.310,-0.282) (-0.288,-0.265) (-0.266,-0.243) (-0.245,-0.225) (-0.223,-0.207) (-0.201,-0.191) (-0.180,-0.177) (-0.158,-0.156) (-0.138,-0.146) (-0.116,-0.124) (-0.095,-0.098) (-0.073,-0.079) (-0.052,-0.057) (-0.032,-0.043) (-0.011,-0.028) (0.011,-0.004) (0.032,0.015) (0.052,0.039) (0.073,0.065) (0.095,0.093) (0.116,0.124) (0.138,0.149) (0.158,0.174) (0.180,0.193) (0.201,0.210) (0.223,0.233) (0.245,0.260) (0.266,0.278) (0.288,0.293) (0.310,0.330) (0.333,0.352) (0.355,0.368) (0.377,0.382) (0.400,0.401) (0.423,0.430) (0.446,0.452) (0.470,0.476) (0.492,0.497) (0.517,0.522) (0.541,0.540) (0.566,0.563) (0.591,0.596) (0.615,0.625) (0.641,0.655) (0.667,0.680) (0.694,0.700) (0.722,0.725) (0.748,0.748) (0.776,0.775) (0.806,0.808) (0.835,0.836) (0.866,0.867) (0.896,0.896) (0.928,0.921) (0.961,0.964) (0.995,1.003) (1.031,1.032) (1.066,1.063) (1.104,1.101) (1.144,1.150) (1.186,1.193) (1.231,1.232) (1.274,1.269) (1.324,1.327) (1.377,1.363) (1.434,1.395) (1.497,1.467) (1.561,1.521) (1.638,1.579) (1.725,1.684) (1.828,1.784) (1.956,1.857) (2.114,2.061) (2.376,2.492) (2.674,2.835) (2.721,2.878) (2.776,2.910) (2.841,2.960) (2.920,3.014) (3.023,3.185) (3.175,3.234) (3.481,3.563)};
\addplot[black, thin, domain=-4.2:4.2] {x};
\end{groupplot}
\end{tikzpicture}
\caption{Normal Q-Q plots of the standardised MST weights, $(M_d-\langle M_d\rangle)/s$, for the surrogate (blue dots, $N=2000$) and the exact Haar model (red open squares, $N=2000$, $2000$ and $500$), with the line $y=x$. About $130$ quantiles per sample, extremes included.}\label{FIG_qq}
\end{figure}
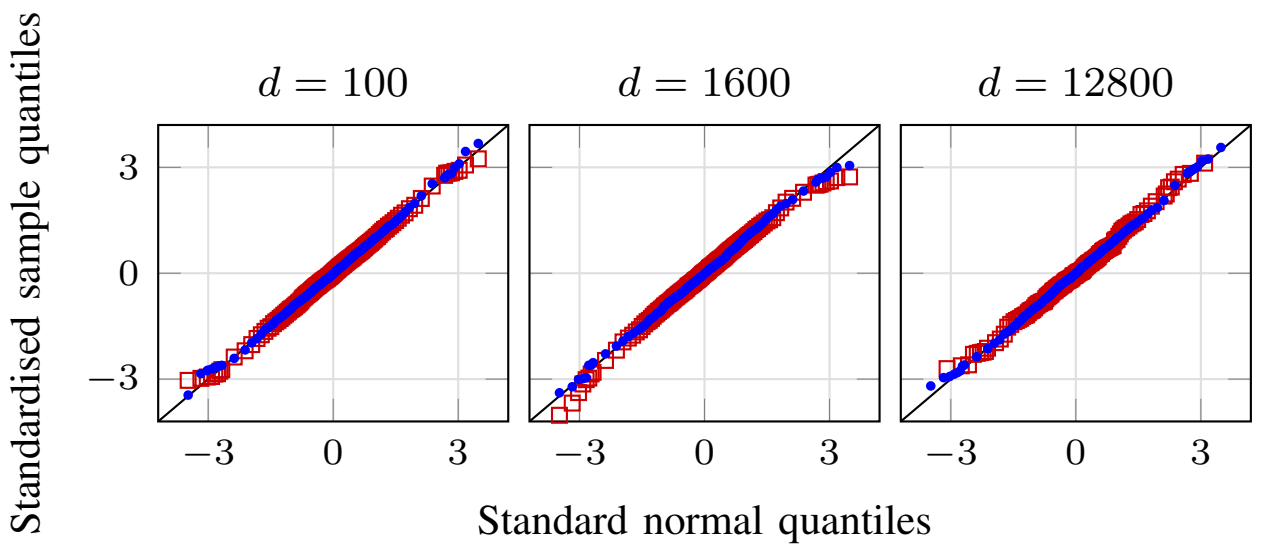

\emph{Fluctuations and leading-order decay.} The sample distributions show no heavy tails. At $d\in\{100,1600,12800\}$, the normal Q-Q plots follow the line $y=x$ up to the extreme tails for both models (Fig.~\ref{FIG_qq}), skewness and excess kurtosis are below $0.15$ in absolute value, and the Shapiro--Wilk test does not reject normality at the $1\%$ level for any of the six samples. Standard errors $s/\sqrt N$ for the means are therefore reliable. Log-log regressions over the full range (Table~\ref{TAB_reg}) give slopes of $\ln\langle M_d\rangle$ versus $\ln\ln d$ equal to $-0.484\pm0.004$ for both models, in agreement with the $(\ln d)^{-1/2}$ envelope of Theorem~\ref{THE_envelope}; the residual deviation from $-1/2$ is the expected signature of the negative $1/\ln d$ correction in \eqref{EQ_expansion}. The means decrease monotonically, which supports Conjecture~\ref{CONJ_general}. The variances, in contrast, decay at different rates, so the Haar constraints do affect the fluctuations, but not the mean at the accessible precision.

\begin{table}[t]
\centering
\caption{Least-squares slopes (95\% confidence intervals) over $d\in[100,18101]$.}\label{TAB_reg}
\footnotesize
\begin{tabular}{llcc}
\toprule
Observable & Model & Slope & $R^2$\\
\midrule
$\ln\langle M_d\rangle$ vs $\ln\ln d$ & Haar & $-0.4839\pm0.0035$ & 0.9998\\
$\ln\langle\Miid\rangle$ vs $\ln\ln d$ & Surrogate & $-0.4838\pm0.0035$ & 0.9998\\
$\ln \mathrm{Var}(M_d)$ vs $\ln d$ & Haar & $-1.3562\pm0.0166$ & 0.9995\\
$\ln \mathrm{Var}(\Miid)$ vs $\ln d$ & Surrogate & $-1.4160\pm0.0263$ & 0.9987\\
\bottomrule
\end{tabular}
\end{table}

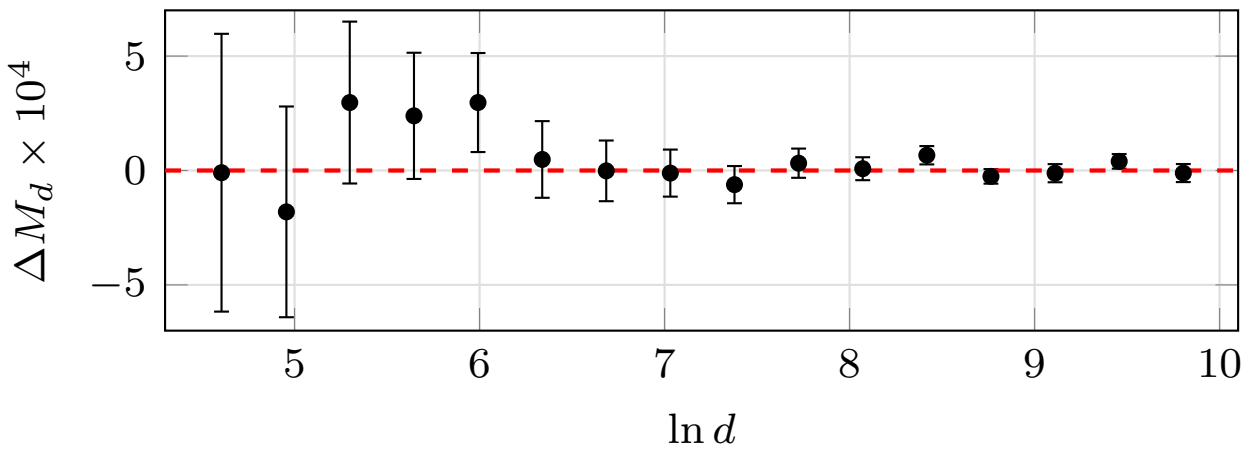
\begin{figure}[t]
\centering
\begin{tikzpicture}
\begin{axis}[
  width=0.98\columnwidth, height=3.7cm,
  xlabel={$\ln d$}, ylabel={$\Delta M_d\times10^{4}$},
  xmin=4.3, xmax=10.1, ymin=-7, ymax=7,
  tick label style={font=\footnotesize}, label style={font=\footnotesize},
  grid=major, grid style={gray!25}]
\addplot[only marks, mark=*, mark size=1.4pt, black,
  error bars/.cd, y dir=both, y explicit] coordinates {
(4.6052,-0.0965) +- (0,6.0706) (4.9558,-1.8079) +- (0,4.6053) (5.2983,2.9699) +- (0,3.5376) (5.6454,2.3889) +- (0,2.7576) (5.9915,2.9705) +- (0,2.1638) (6.3386,0.4826) +- (0,1.6752) (6.6846,-0.0157) +- (0,1.3264) (7.0309,-0.1138) +- (0,1.0285) (7.3778,-0.6183) +- (0,0.8127) (7.7244,0.3195) +- (0,0.6385) (8.0709,0.0780) +- (0,0.5012) (8.4174,0.6668) +- (0,0.4005) (8.7641,-0.2569) +- (0,0.3181) (9.1106,-0.1178) +- (0,0.3965) (9.4572,0.3986) +- (0,0.3190) (9.8037,-0.1119) +- (0,0.3942)};
\addplot[red, dashed, thick, domain=4.3:10.1] {0};
\end{axis}
\end{tikzpicture}
\caption{Residual $\Delta M_d=M_d-\Miid$ between the exact Haar model and the surrogate, with error bars $\pm\sigma_\Delta$ (one standard error).}\label{FIG_delta}
\end{figure}

\emph{Consistency of the exact model with the surrogate.} The residual $\Delta M_d$ has standard error $\sigma_\Delta=(s_{\mathrm{H}}^2/N_{\mathrm{H}}+s_{\mathrm{iid}}^2/N_{\mathrm{iid}})^{1/2}$. Over the sixteen dimensions, $|\Delta M_d|\le3.0\times10^{-4}$ (reached at $d=400$), $|\mathrm{SNR}|\le1.66$ (reached at $d=4525$), the sign varies without pattern, and no trend in $d$ is visible (Fig.~\ref{FIG_delta}). The sum of the squared SNRs equals $9.6$ for $16$ degrees of freedom ($p\approx0.89$ under the null hypothesis of equal means), so the two models are statistically indistinguishable in mean over the whole range, with a resolution of a few $10^{-5}$ at the largest dimensions.

\begin{figure}[t]
\centering
\begin{tikzpicture}
\begin{axis}[
  width=0.98\columnwidth, height=4.5cm,
  xlabel={$x(d)=1/\ln d$}, ylabel={$y(d)=M_d\sqrt{\ln d}$},
  xmin=0, xmax=0.225, ymin=3.255, ymax=3.34,
  tick label style={font=\footnotesize}, label style={font=\footnotesize},
  xtick={0,0.05,0.10,0.15,0.20},
  xticklabel style={/pgf/number format/fixed,/pgf/number format/precision=2,/pgf/number format/fixed zerofill},
  yticklabel style={/pgf/number format/fixed,/pgf/number format/precision=2,/pgf/number format/fixed zerofill},
  legend cell align=left, legend columns=2,
  legend style={font=\scriptsize, at={(0.02,0.04)}, anchor=south west, draw=none, fill=white, fill opacity=0.85, text opacity=1,
    inner sep=1pt, row sep=-1pt, /tikz/every even column/.append style={column sep=6pt}},
  legend image post style={scale=0.7},
  grid=major, grid style={gray!25}]
\addplot[only marks, mark=*, mark size=1.5pt, blue,
  error bars/.cd, y dir=both, y explicit] coordinates {
(0.21715,3.26227) +- (0,0.00195) (0.20178,3.27507) +- (0,0.00153) (0.18874,3.28237) +- (0,0.00119) (0.17713,3.28956) +- (0,0.00094) (0.16690,3.29366) +- (0,0.00074) (0.15776,3.29806) +- (0,0.00059) (0.14960,3.30025) +- (0,0.00049) (0.14223,3.30213) +- (0,0.00037) (0.13554,3.30385) +- (0,0.00031) (0.12946,3.30499) +- (0,0.00025) (0.12390,3.30587) +- (0,0.00020) (0.11880,3.30663) +- (0,0.00016) (0.11410,3.30748) +- (0,0.00013) (0.10976,3.30799) +- (0,0.00011) (0.10574,3.30860) +- (0,0.00009) (0.10200,3.30913) +- (0,0.00007)};
\addlegendentry{Surrogate}
\addplot[only marks, mark=square, mark size=2.0pt, red,
  error bars/.cd, y dir=both, y explicit] coordinates {
(0.21715,3.26225) +- (0,0.00165) (0.20178,3.27467) +- (0,0.00130) (0.18874,3.28305) +- (0,0.00106) (0.17713,3.29012) +- (0,0.00087) (0.16690,3.29439) +- (0,0.00073) (0.15776,3.29819) +- (0,0.00058) (0.14960,3.30024) +- (0,0.00046) (0.14223,3.30210) +- (0,0.00038) (0.13554,3.30368) +- (0,0.00030) (0.12946,3.30508) +- (0,0.00024) (0.12390,3.30590) +- (0,0.00019) (0.11880,3.30682) +- (0,0.00016) (0.11410,3.30741) +- (0,0.00013) (0.10976,3.30796) +- (0,0.00021) (0.10574,3.30872) +- (0,0.00017) (0.10200,3.30910) +- (0,0.00023)};
\addlegendentry{Exact (Haar)}
\addplot[black, thick, smooth] coordinates {
(0.0000,3.33333) (0.0100,3.32904) (0.0200,3.32534) (0.0300,3.32216) (0.0400,3.31945) (0.0500,3.31715) (0.0600,3.31518) (0.0700,3.31349) (0.0800,3.31201) (0.0900,3.31068) (0.1000,3.30936) (0.1100,3.30803) (0.1200,3.30654) (0.1300,3.30480) (0.1400,3.30267) (0.1500,3.30005) (0.1600,3.29682) (0.1700,3.29288) (0.1800,3.28814) (0.1900,3.28248) (0.2000,3.27584) (0.2100,3.26822) (0.2200,3.25942)};
\addlegendentry{Constrained fit}
\addplot[gray, dashed, domain=0:0.225] {10/3};
\addlegendentry{$10/3$}
\end{axis}
\end{tikzpicture}
\caption{Renormalised MST weight versus the finite-size variable $x(d)=1/\ln d$, with 95\% confidence intervals (mostly smaller than the markers). The solid line is the constrained fit of Table~\ref{TAB_fits} ($d_{\min}=100$), extrapolated to $x=0$.}\label{FIG_y}
\end{figure}
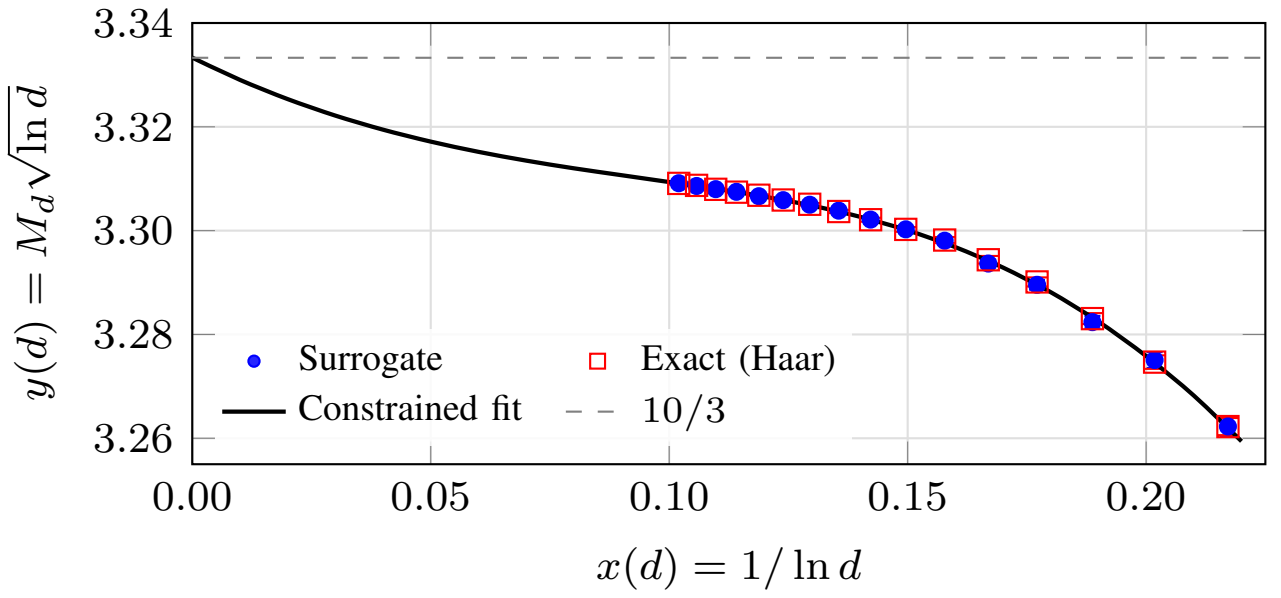

\emph{Thermodynamic limit.} We test \eqref{EQ_y} in two ways, by weighted least squares (WLS) with weights given by the inverse variances of the sample means of $y(d)$ (Table~\ref{TAB_fits}). First, a free quadratic extrapolation $y(d)=\alpha+\beta x+\gamma x^2$, $x=1/\ln d$, applied to the dimensions $d\ge d_{\min}$: the intercept increases with $d_{\min}$ for both models and stabilises around $3.317$, i.e., $0.5\%$ below $10/3$, while $\chi^2/\nu$ decreases from about $8$ to about $1$. Second, a fit constrained by the theory: imposing the limit, i.e., fitting $r(d)=y(d)\sigma_d/d$ as $r=1+Ax+Bx^2+Cx^3$, gives $\chi^2/\nu\approx1$ over the \emph{full} range for both models. The data are thus fully consistent with $\alpha=10/3$; the offset of the free extrapolation reflects the curvature of the logarithmic corrections. Fig.~\ref{FIG_y} shows that all samples of $y(d)$ lie below $10/3$ and increase towards it, for both models.

\begin{table}[t]
\centering
\caption{WLS fits of $y(d)=M_d\sqrt{\ln d}$ on $d\ge d_{\min}$ (one standard error in parentheses, in units of the last digit). Free: $y=\alpha+\beta x+\gamma x^2$. Constrained: $y\,\sigma_d/d=1+Ax+Bx^2+Cx^3$, with $x=1/\ln d$.}\label{TAB_fits}
\scriptsize
\setlength{\tabcolsep}{2.2pt}
\begin{tabular}{rcccc}
\toprule
 & \multicolumn{2}{c}{Free: $\alpha$ ($\chi^2/\nu$)} & \multicolumn{2}{c}{Constrained: $A$ ($\chi^2/\nu$)}\\
\cmidrule(lr){2-3}\cmidrule(lr){4-5}
$d_{\min}$ & Surrogate & Haar & Surrogate & Haar\\
\midrule
100  & 3.2924(10) (9.2) & 3.2868(12) (7.3) & $-0.138(2)$ (1.0) & $-0.140(3)$ (0.7)\\
566  & 3.3080(19) (1.5) & 3.3046(26) (1.1) & $-0.131(5)$ (0.6) & $-0.133(6)$ (0.9)\\
1600 & 3.3167(35) (0.6) & 3.3111(51) (1.2) & $-0.122(9)$ (0.6) & $-0.134(13)$ (1.0)\\
3200 & 3.3179(66) (0.9) & 3.317(10) (1.3) & $-0.125(18)$ (0.9) & $-0.125(27)$ (1.2)\\
\bottomrule
\end{tabular}
\end{table}

\emph{Subleading coefficient.} A one-parameter fit $r(d)=1+A/\ln d$ gives the effective value $A\approx-0.069$, stable in $d_{\min}$; with an additional $1/\ln^2 d$ term, $A\approx-0.097$ ($d_{\min}=1600$, $\chi^2/\nu=1.7$); with the cubic of Table~\ref{TAB_fits}, $A\approx-0.12$ to $-0.14$. These values bracket the asymptotic prediction $A\approx-0.111$, and their spread shows that $A$ is strongly correlated with the higher-order coefficients: at $d=18101$, $|A|/\ln d\approx0.011$ is comparable to the next term. All estimates agree on the sign $A<0$, which fixes the approach to $10/3$ from below.

\section{Conclusion}

For single-input bilinear quantum systems with drift spectrum $k^{3/2}$ and Haar-random control eigenbasis, the MST weight that bounds the state-to-state control time through \eqref{EQ_upper} behaves, at all tested dimensions $100\le d\le18101$, exactly as in the independent Gaussian surrogate, for which we derived $M_d\asymp(\ln d)^{-1/2}$ and the limit $M_d\sqrt{\ln d}\to10/3$. This supports Conjectures~\ref{CONJ_general}--\ref{CONJ_precise} with $\alpha=10/3$, and therefore $\tau_{SS}^{(d)}\lesssim\frac{5\pi}{3}(\ln d)^{-1/2}$ for generic systems: at the critical growth, the minimal control time vanishes, but only logarithmically. From a control-design viewpoint, the critical exponent $3/2$ thus separates spectra for which adding levels accelerates generic state transfer polynomially from spectra for which the gain is marginal. A proof for the exact Haar model, which requires controlling the global correlations beyond finite clusters, and the analysis of the lower bound of Proposition~\ref{PRO_bounds} are left for future work.

\section*{Acknowledgment}
The authors also acknowledge the use of Gemini and the large‑language‑model services
provided by the University of Strasbourg (Unistra) for writing the Python code, drafting the manuscript, generating the pictures and converting
the original French notes into academic English.  While these tools contributed to the writing, the authors
take full responsibility for the scientific integrity, mathematical correctness, and
final verification of this note.

\bibliographystyle{IEEEtran}
\bibliography{lcss_biblio}

\end{document}